\documentclass[ amsmath,amssymb, aps,prl, reprint]{revtex4-2}

\usepackage[utf8]{inputenc}
\usepackage{graphicx}
\usepackage{hyperref}
\usepackage{aas_macros}
\usepackage{makecell}
\usepackage{xcolor}
\usepackage{comment}
\begin{document}

\title{Optical ultra-stable optical clock cavities as  resonant mass  gravitational wave detectors in search  for new physics.}
\author{M. Naro\.znik, M. Bober, M. Zawada}

%\date{\today}
\address{Institute of Physics, Faculty of Physics, Astronomy and Informatics, Nicolaus Copernicus University, Grudzi\c{a}dzka 5, PL-87-100 Toru\'n, Poland}

\begin{abstract}
We propose to use table-top-size ultra-stable optical cavities from the state-of-the-art optical atomic clocks  as  bar gravitational wave detectors for the frequencies higher than 2~kHz. We show that 2-20~kHz range of gravitational waves' spectrum can be accessed with instruments below 2~meters in size. The proposed cavities’ materials and properties are being within the present-day technology grasp. The ultra-stable optical cavities allow detecting not only  predicted gravitational wave signals from such sources as binary neutron star mergers and post-mergers, subsolar-mass primordial black-hole mergers, and collapsing
stellar cores, but can reach new physics beyond standard model looking for ultralight bosons such as QCD axions and axion-like particles formed through black hole superradiance.

\end{abstract}

\maketitle

\section{Introduction} 

The first generation of gravitational waves (GWs) detectors was based
on mechanical resonance in large aluminium cylinders that is triggered by gravitational wave flying through the detector \cite{Weber0}. 
This type of GWs detectors, so-called Weber bars, are still operating connected in the worldwide network. 
Unfortunately, despite a superb stable operation and a perfect reduction of noises in the present-day resonant-mass bar detectors,
so far there is no evidence of GWs events observed by the acoustic resonance phenomena \cite{Abbot_2007}.

The second type of detectors is based on Michelson interferometry. 
In such detectors ripples in the space-time are registered by electromagnetic radiation instead of acoustic waves. 
In 2015 the ground-based observatories were finally sensitive enough to detect the first gravitational wave event, which was two black holes (BHs) merger \cite{GW_detection1}.
{So far, during three observational runs 90 events were confirmed in total, where vast majority were BHs  merger (BH-BH), two of black hole~-~neutron star merger and two neutron~-~neutron star (NS-NS) mergers  \cite{Poggiani_2021_03results}}
In the events observed up to date the total mass and luminosity distance (i.e., distance from the Earth calculated by measuring absolute and apparent magnitudes) of binary systems were in the range  from  $2.74^{+0.04}_{-0.01}$ solar masses (M$_{\odot})$ and $40^{+8}_{-14}$~Mpc (NS-NS) to 
$142^{+28}_{-16}$~M$_{\odot}$ and  $5.3^{+2.4}_{-2.6}$~Gpc, which is the first observational evidence of the existence of a BH with a mass of more than 100~M$_{\odot}$, so-called intermediate-mass black hole) \cite{Abbott_2020_IMBH}.

The range of maximum sensitivity for existing  LIGO-like observatories spans between $\sim10$~Hz and roughly 2~kHz~\cite{Sensitivity_2018}.
Most of the other promising experiments, like the International Pulsar Timing Array (IPTA)~\cite{Hobbs_2010} that includes three independent projects, i.e. NANOGrav~\cite{NANOGrav_2018},  EPTA~\cite{EPTA_2015}, and PPTA~\cite{PPTA_2015}, 
optical atomic clocks~\cite{Kolkowitz_2016}, torsion-bar antenna TOBA~\cite{Ando_2010},   atomic interferometry projects, like AION  \cite{AION_2020} and ELGAR  \cite{ELGAR_2019}, and the space-based LISA~\cite{Amaro_2017} and DECIGO \cite{Sato_2017}
are sensitive at frequencies lower than existing LIGO-like detectors.

At present there is a noticeable gap in higher frequency range, although several important GWs sources probably emit in this  part of the GW spectrum, and, moreover,  GWs detectors sensitive in 1-100~kHz range are perfectly suited for searches beyond the standard model.
Recently proposed  levitated-sensor-based GW detector~\cite{Aggarwal_2022}, that would achieve reasonable sensitivity in the above 10~kHz range, requires cryogenic 100~m facility. Underground LIGO-like Einstein Telescope~\cite{ET_2020}, being still in the early design study phase, will have 10 km long arms.

In this letter we prove that table-top-size ultra-stable optical cavities from the state-of-the-art optical atomic clocks~\cite{Ludlow_2015,OC18} can be used as  resonant-mass gravitational wave
detectors for the frequencies higher than 2~kHz. We calculate the mechanical resonances  for the  existing state-of the art and future possible cavity set-ups and 
analyse limitation of sensitivity  by fundamental noises. 
The proposed cavities' materials and components were selected for the best properties while being within the present-day technology grasp. We show that fundamentally limited sensitivity to GW of a table-top ultra-stable optical cavity used as a resonant-mass gravitational wave
detector allows detecting predicted GW signals from such sources as  binary neutron star mergers and post-mergers, collapsing stellar cores, subsolar-mass primordial BHs  mergers, and QCD axions and axion-like particles formed through BH superradiance.

\section{Principle of observation}

A behaviour of a bar-like resonant detector, e.g. used in the Weber resonant mass GWs detectors, with resonant frequency of $f_0$,
in the vicinity of the gravitational wave can be modelled as a driven and damped harmonic oscillator~\cite{Creighton} in a form of two masses 
connected by a spring of a length $L$.
The  spring constant $k$ and  the masses  are chosen to satisfy $f_0=1/2\pi\sqrt{k/\mu}$, where $\mu$ is the reduced mass of the system. 
 
The mass displacement $x$ induced by a plane gravitational wave with frequency $f$,  travelling perpendicularly to the spring and polarised along the spring, can be described by a simple equation of motion

\begin{equation}
    \ddot{x}(t) + 2\beta \dot{x}(t)  + 4\pi^{2}f_{0}^{2}x(t) = F_{GW}(t), 
    \label{eq:oscillator}
\end{equation}

\noindent 
where $\beta := \pi f_{0}/ Q$ is a damping parameter related by definition with a  quality factor of the spring $Q$,  and $F_{GW}=-\frac{1}{2}hL(2\pi f)^{2}\cos(2\pi f t)$ is a GW force acting on the system. The strain amplitude of the GWs is denoted as $h$. 

The eq. \ref{eq:oscillator} can be  solved in the frequency domain, with corresponding quantities $\tilde{x}(f) = G(f) \tilde{h}(f)$, where $\tilde{h}(f)$ is a strain in the frequency domain and $ G(f)$ is the following  transfer function

\begin{equation}
    G(f) = \frac{L}{2}\frac{f^2}{(f_{0}^2 - f^2) + iff_{0}/Q }.
\end{equation}

The best sensitivity to the GW that can be achieved by the system, i.e.
the GW strain-equivalent power spectral density (PSD)   $S_{h}(f)$, is related to the PSD  of the noise present in the system $S_{x}$ by

\begin{equation}
    S_{h}(f) = \frac{S_{x}(f)}{|G(f)|^{2}}.
\end{equation}

In general, 
$f_{0}$ is a characteristic resonance frequency of a detector. A simple three-dimensional analytical approach shows that position of the $n_l$-th resonance 
$f_{0} = \frac{1}{2L} \sqrt{\frac{E}{\rho}n_{l}^2}$ 
depends on detector length $L$ and material internal properties i.e. Young modulus $E$ and density $\rho$. 

\begin{figure}[hbt]
    \centering
    \includegraphics[width=0.7\columnwidth]{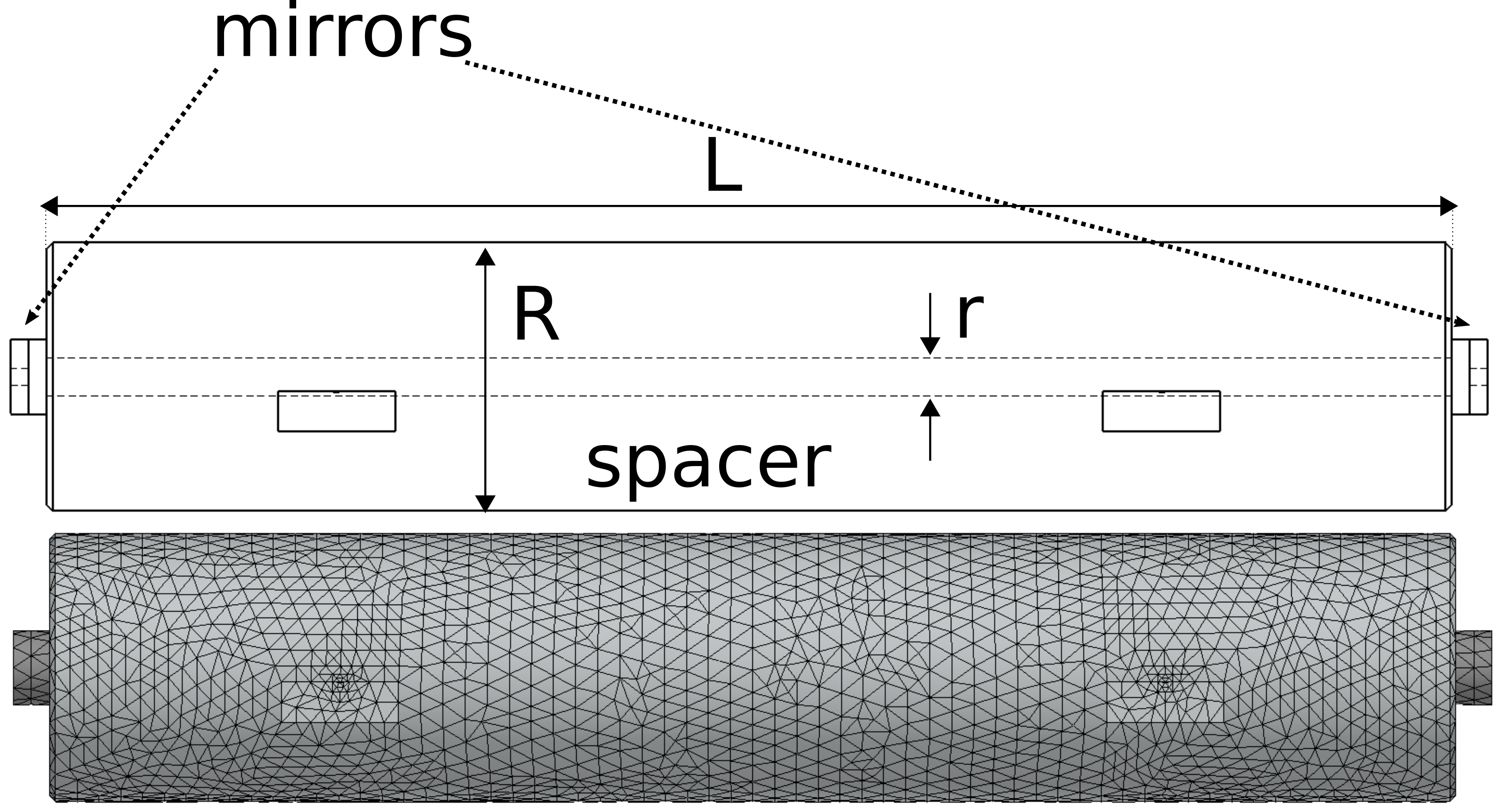}
    \caption{A cross section of a typical ultra-stable optical cavity used in optical atomic clocks experiments and mesh for the FEM simulation. $L$ and $R$ stands for its length and external radius, respectively, while $r$ is the radius of internal bore between mirrors. }
    \label{fig:FEM}
\end{figure}

In the case of an ultra-stable optical cavity this simplified approximation is not  sufficiently accurate in determining exact values of the cavity resonance.
Moreover, in existing system the optical cavities are placed on carefully calculated points to dump most of the possible mechanical resonances. 
Therefore, we performed a finite element method (FEM) simulation for the several existing state-of the art and future possible cavity set-ups, taking into account Earth gravity field and support points. Fig.~\ref{fig:FEM} shows a cross section of a popular horizontal design of a cavity used the optical clock experiments together with the mesh used in our FEM simulation. The resolution of the mesh is adjusted to  the local radius of curvature in the range of {1.5~mm to 7~mm}. 
Fig.~\ref{fig:FEM} also defines geometrical variables of the spacer, i.e. spacer radius $R$ and length $L$, and internal bore diameter $r$.
In the case of the spacer made of ultra-low expansion (ULE) glass and mirrors' substrate made of fused-silica (FS), additional ULE rings are usually added outside the mirrors to tune  the zero crossing temperature of the cavity~\cite{Lagero_2010} - these rings, however, do not contribute significantly to the mechanical resonance properties of the whole system.

\begin{figure}[hbt]
    \centering
    \includegraphics[width=1\columnwidth]{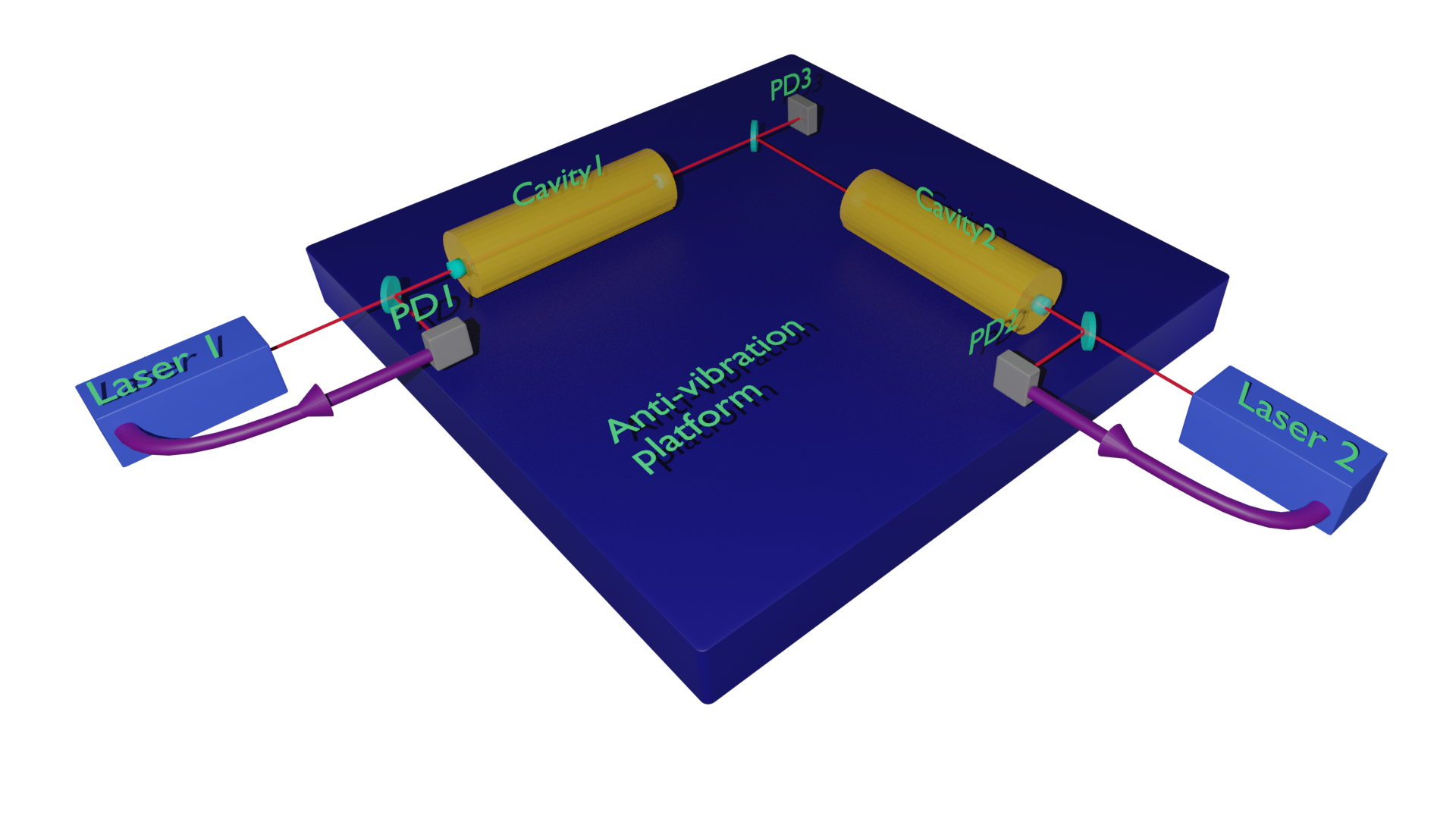}
    \caption{(colour online) The two cavities are aligned perpendicularly to each other. The relative length change between cavities is detected by frequency or phase difference between the lasers' beams stabilised to the cavities. PD stands for photodiode.}
    \label{fig:setup}
\end{figure}

To detect the  length change of an optical cavity a laser light is frequency-locked to one of the cavity modes.
Shrinking and extending path of a photon inside the optical cavity will effectively move the cavity mode frequency $\nu$ by $\Delta\nu$ according to the simple formula ${\Delta L}/{L} = -{\Delta \nu}/{\nu} $. With a laser frequency tightly stabilised to the cavity mode frequency, e.g. by the Pound-Drever-Hall technique~\cite{Black_2001},  the length change of the cavity is transferred to the phase or frequency change of the laser light.

To detect change of the length of a cavity due to GW, a reference is needed, for instance a second, perpendicular cavity. Fig.~\ref{fig:setup} depicts a system of two ultra-stable optical cavities aligned perpendicularly to each other, either in the horizontal or in the vertical plane.  Environmental perturbations can be greatly reduced by installing both cavities in one shared vacuum system and mounting the system on a single vibration isolation platform. With two lasers’ beams stabilised to each cavity, the relative length change between the cavities is transferred to the phase or frequency difference between the lasers’ light. This difference can be detected by an optical beat note on a photodiode.
The in-vacuum detection of beat note signal
may by done either with the light transferred through both cavities (like in Fig.~\ref{fig:setup}), or by with light uncoupled from main beams in front of cavities. In the former case, the light is filtered by the optical cavities but has low intensity, while in the latter case the light can has high intensity improving the beat note detection signal.
The beat note detection system as well as optical elements and photodiodes required for lasers stabilisation can be placed in-vacuum on the same vibration isolation platform. The light from the lasers can be transferred to the platform over fibres. With such configuration the cavities can be surrounded in vacuum by several thermal shields providing superb thermal insulation. Additionally, installing the cavities and the beat note system in the mutual vacuum set-up allows skipping the  optical path length stabilisation~\cite{Falke_2012}.

\section{Fundamental sensitivity limits}

The fundamental sensitivity of an ultra-stable optical cavity to a GW is primarily limited by thermal processes i.e. mechanical and optical thermal noise. {The mechanical thermal noise }refers to the Brownian motion of a cavity residing in non-zero temperature $T$. The thermodynamical fluctuations of the cavity components can be expressed quantitatively by the fractional PSDs $S_{y,t}$ 
using fluctuation-dissipation theorem~\cite{Callen_1951, Callen_1952}. The {magnitude of} mechanical thermal noise depends on the 
spacer geometry 
and mass $m$, and {mirrors'} coating thickness $d_{ct}$, {as well as on the} cavity intrinsic physical parameters, such as Young modulus $E$, Poisson's coefficient $\sigma$, and mechanical loss angle $\phi$:

\begin{widetext}

\begin{equation}
    S_{y,t}(f) = \frac{4k_{b} T \phi_{sp}}{(2\pi)^{3}mL^{2}} \frac{f_{0}^2}{f[(f_{0}^2-f^{2})^{2} + f_{0}^{4}\phi_{sp}^{2}]} + 
    \frac{4k_{b}T}{2\pi^{5/2}}\frac{1-\sigma_{sb}^{2}}{f E_{sb} w L^{2} }\phi_{sb} 
  \left(
  1+\frac{2d_{ct}}{w\sqrt{\pi}}\frac{1-2\sigma_{sb}}{1-\sigma_{sb}}\frac{\phi_{ct}}{\phi_{sb}}
 \right).
 \label{eq:ASD_thermal}
\end{equation}
\end{widetext}

\noindent
where $sp$, $sb$, $ct$ indices corresponds to spacer, mirrors' substrate, and mirrors' coating, respectively, and $w$ is the beam spot radius on the mirror. Fractional PSD $S_{y}(f)$ is related to the PSD of the length fluctuation $S_{x}(f)$ by $S_{y}(f) = S_{x}(f)/L^{2}$.

Additionally, the contribution to the total  fractional PSD from the non-Brownian optical thermal noise composed of thermo-elastic and thermo-refractive noises in mirror substrate and multilayer mirror coatings can be described by fractional PSD $S_{y,o} \sim L^{-2} 
    (1+\sigma_{sb})^2\alpha^{2}_{sb}{T^{2}}A(w,f) + L^{-2} {T^{2}}B(w,f)$,
where $\alpha_{sb}$ is the mirror substrate coefficient of thermal expansion and  $A(w, f)$ and $B(w,f)$ are effective material constants of mirrors' substrate and coatings (see ~\cite{Cole_2013}).

In practical realisations, intensity of light used to measure the cavity length is kept at minimum detectable levels to avoid additional heating of the mirrors surface. This leads to another potential source of sensitivity limitation due to the quantum nature of light, the shot noise $S_{y,s} =\sqrt{2\pi\hbar c}/(8\mathcal{F}L\sqrt{\lambda P_{c}})$, where $\mathcal{F}$ is the optical cavity finesse, $\hbar$ is the reduced Planck constant, $\lambda$ is the light wavelength, and $P_{c}$ is the power of light injected into the cavity~\cite{Black_2001}. This noise, however, will be further minimised as newer mirror coatings materials will allow for higher intra-cavity light powers.

\begin{figure}[hbt]
    \centering
    \includegraphics[width=\columnwidth]{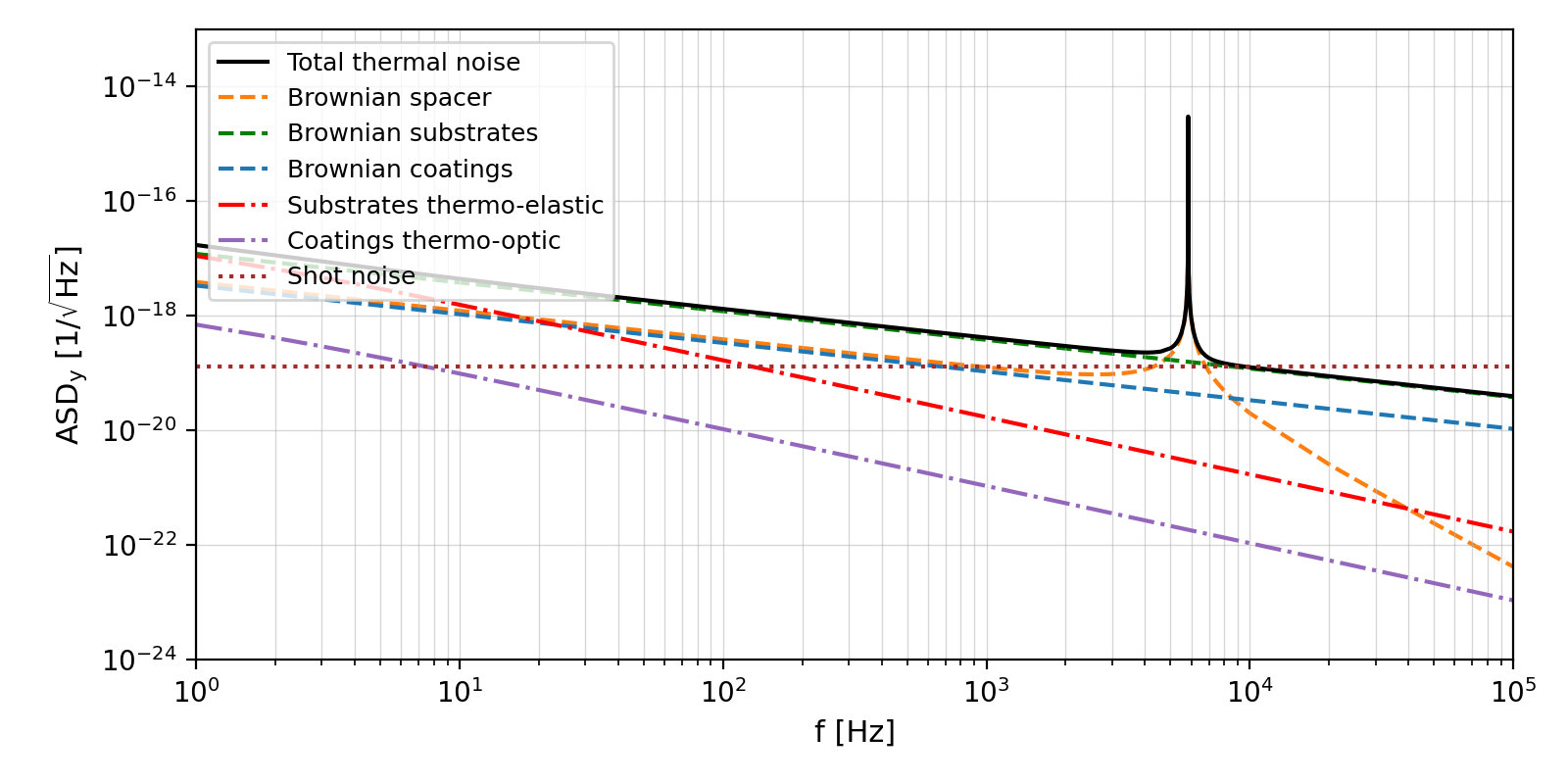}\quad
    \includegraphics[width=\columnwidth]{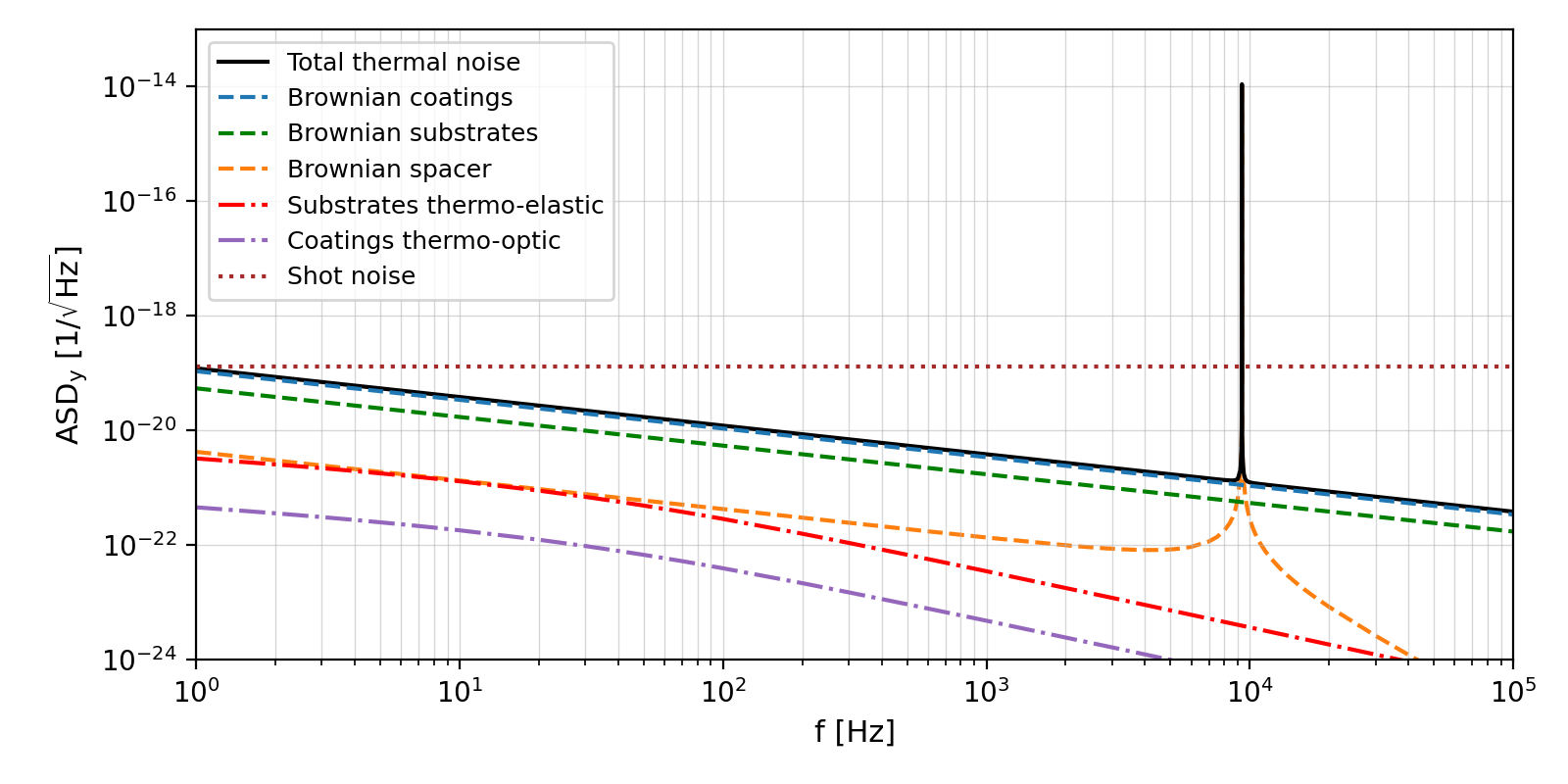}\quad
    \includegraphics[width=\columnwidth]{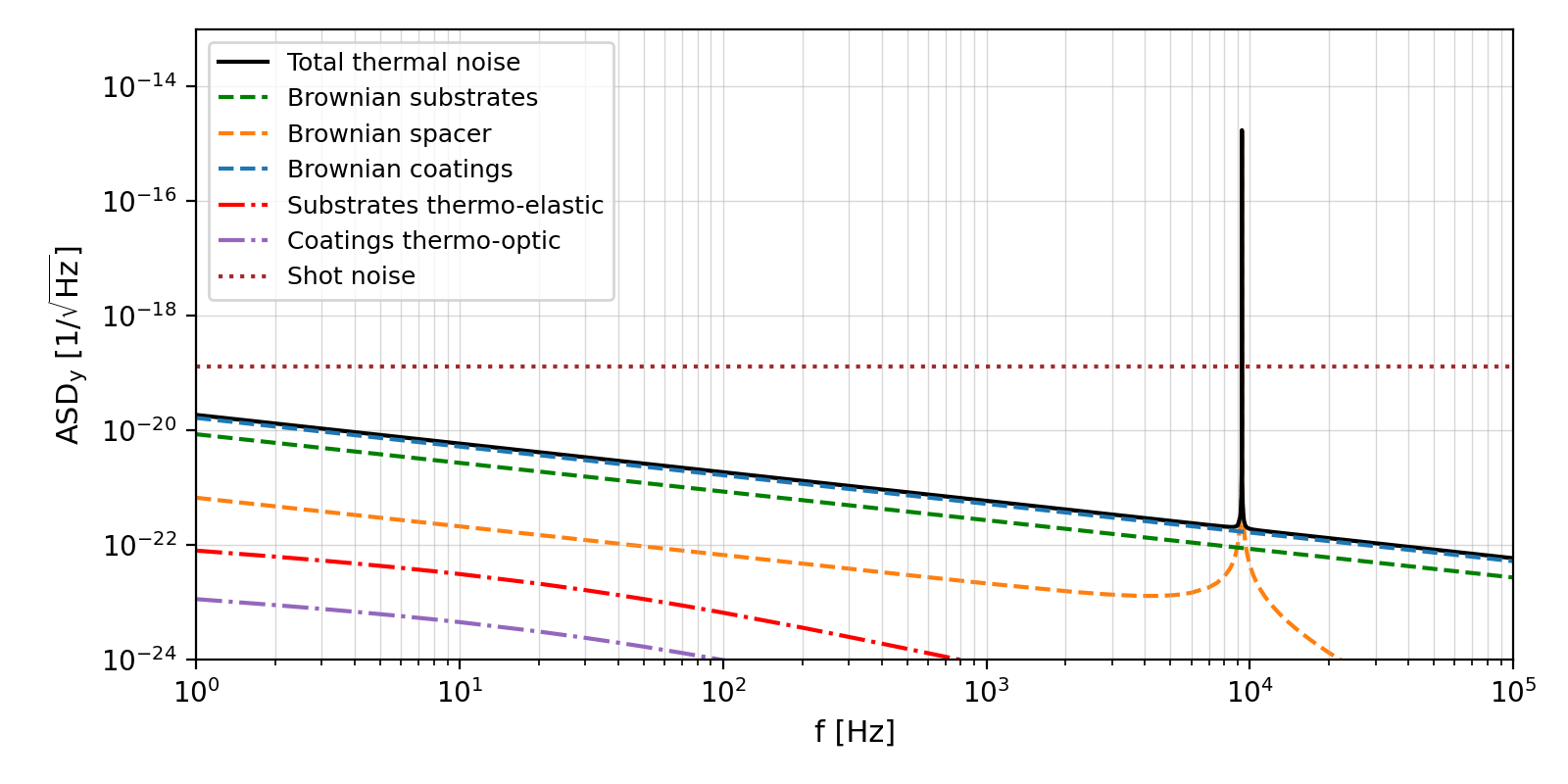}\quad
    \caption{(colour online) {Estimated sensitivities limited by fundamental noises represented by amplitude spectral density (ASD)} of  a present-generation ultra-stable optical cavity  (0.5~m long cavity made of ultra-low expansion (ULE) glass at room temperature) if the mirrors' coatings would be replaced by crystalline AlGaAs coatings~\cite{Hafner_2015} (top),  possible cryogenic 0.5-m long cavity made of  single-crystal silicon with  mirrors with crystalline AlGaAs coatings at 4~K (middle) and  in 0.1~K (bottom). The exemplary shot-noise limit is estimated for 50~$\mu$W cavity light power  and  the cavity finesse of {150000}.}
    \label{fig:noises}
\end{figure}

\begin{figure}
    \centering
    \includegraphics[width=1\columnwidth]{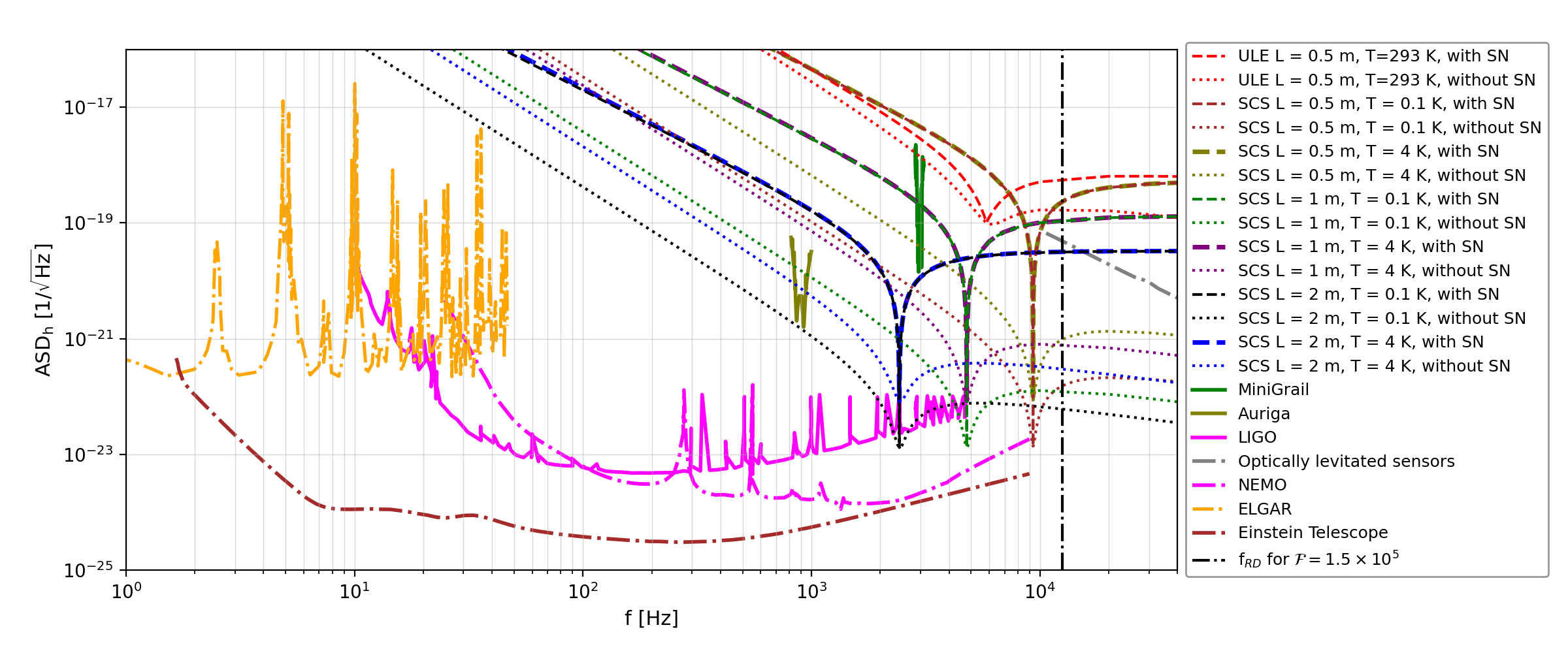}
    \caption{Estimated sensitivities to GWs of the existing 0.5~m long optical cavity made of ULE glass at room temperature if the mirrors' coatings would be replaced by crystalline AlGaAs coatings~\cite{Hafner_2015} (red) and possible cryogenic 0.5~m, 1~m and 2-m long cavities made of  single-crystal silicon (SCS) at 4~K and 0.1~K.  All cases have been presented without (dashed lines) the shot-noise (SN), which corresponds to the hard fundamental limits, and with (dotted lines) the exemplary shot-noise limit estimated for 50~$\mu$W  light power injected into the cavity of {150000} finesse. The sensitivities of other existing~\cite{LIGO_1st_volume_1989, Auriga_2016, MiniGrail_2007}
    (solid lines) and planned~ \cite{Ackley_2020_NEMO,ELGAR_2019,Aggarwal_2022,Moore_2015} (dash-dot lines) GWs detectors are added to the plot for comparison. Vertical black dot-dashed line represents exemplary ring-down frequency limit for the 0.5~m long cavity with finesse $\mathcal{F} = 150000$. }
    \label{fig:sensitivity}
\end{figure}

{Fig.~\ref{fig:noises} shows
estimated sensitivities limited by fundamental noises represented by fractional amplitude spectral density 
$A_{y} = \sqrt{S_{y}}$  (where $S_{y}$ includes all previously described noise components)
of the existing  state-of-the-art ultra-stable optical cavity  (0.5~m long cavity made of ULE glass at room temperature) if the mirrors' coatings would be replaced by crystalline AlGaAs coatings~\cite{Hafner_2015} and possible cryogenic 0.5~m and 1-m long cavities made of  single-crystal silicon  with mirrors with crystalline AlGaAs coatings at 4~K and 0.1~K.}
All cases have been presented without the shot-noise, which corresponds to the hard fundamental limits, and with the exemplary shot-noise limit estimated for 50~$\mu$W  light power injected into the cavity  of finesse $\mathcal{F} = 150000$. 

In Fig.~\ref{fig:sensitivity} we depict
estimated amplitude strain sensitivities $A_{h}(f) =\sqrt{S_{h}(f)}$ to GWs of several possible ultra-stable optical cavities. The sensitivities of other existing (LIGO~\cite{LIGO_1st_volume_1989}, AURIGA~\cite{Auriga_2016} and MiniGrail~\cite{MiniGrail_2007}) and planned (NEMO~\cite{Ackley_2020_NEMO}, ELGAR~\cite{ELGAR_2019}, Optically levitated sensors~\cite{Arvanitaki_2013,Aggarwal_2022}, and Einstein telescope\cite{Moore_2015}) GWs detectors are added to the plot for comparison.

\section{Discussions}

The detectable range of the Fabry-Per\'ot detector is also limited in frequency spectrum, because of the finite speed of light and the reflection of the mirrors surface. 
The finesse is responsible for the number of reflections of a photon inside the optical cavity before it leaves it and can contribute to the feedback to the laser frequency --- the higher finesse of the cavity, the longer ring-downtime for a photon in the cavity and the lower servo loop bandwidth. The highest GWs frequency that can be detected for a given~$\mathcal{F}$ is limited by $f_{RD}<\frac{\pi c }{\mathcal{F} L}$. For the 0.5~m long cavity with finesse  $\mathcal{F} = 150000$ the limit is $f_{RD} \sim {12566}$~Hz.

\begin{figure}
    \centering
    \includegraphics[width=1\columnwidth]{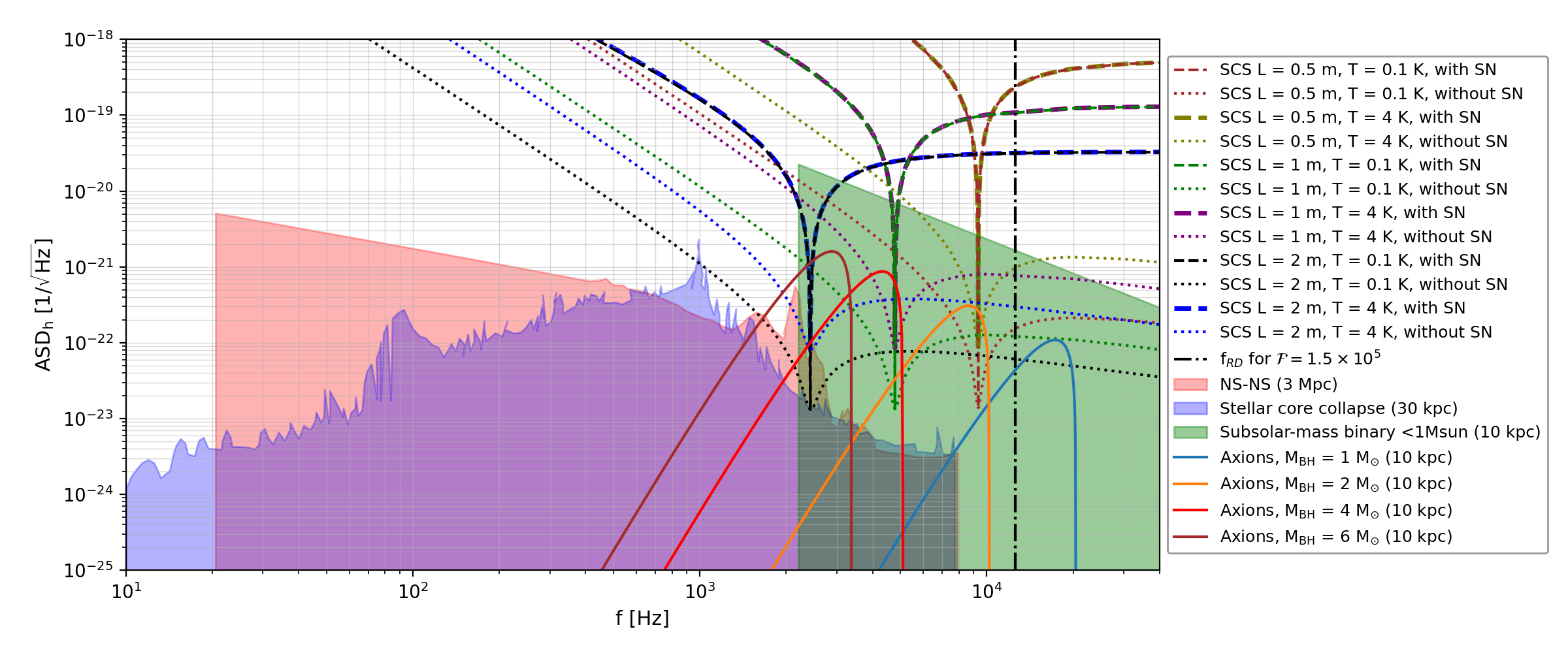}
    \caption{ Comparison of GWs sensitivity limits of technically possible optical cavities with predicted GWs signals of several possible sources. Dotted and dashed lines depict sensitivities to GWs signals 0.5~m, 1~m, and 2~m  long cavities made of SCS at 4K. All cases have been presented without (dashed lines) the shot-noise (SN), which corresponds to the hard fundamental limits, and with (dotted lines) the exemplary shot-noise limit
estimated for 50~$\mu$W light power injected into the cavity of 150000 finesse. Vertical black dot-dashed line represents exemplary ring-down frequency limit for the 0.5 m long cavity with finesse F = 150000. 
The brick-red area depicts the GWs signal of a typical binary neutron star (NS) inspiral, merger, and post-merger  (NS-NS)~\cite{Ackley_2020_NEMO} at the distance of 3 Mpc.
The blue area presents the characteristic GWs spectra from  the process of a BH formation from fast spinning, 
moderate-metallicity, massive stellar progenitors~\cite{Cerda_2013_collapse}  at
the distance of 30 kpc.
Potential GWs signals from the primordial sub-solar mass BHs, calculated analytically for the innermost stable circular orbits of the BHs binaries with two equal masses, each $<$1~M$_{\odot}$, at the distance of   1~kpc \cite{Creighton}, 
were depicted by green area.
The solid lines present predicted signals due to GWs emitted
from axions or ALPs around BHs  in our galaxy within 10 kpc for
$10^6$s coherent integration time. The BH with initial masses of 1, 2, 4 and 6 M$_{\odot}$ (blue, orange, red and brown, respectively) and initial spin of 0.9 were calculated for the dominant level ($l = m_l = 1$, $n = 0$)~\cite{Isi_2019,Isi_2020}.
}
    \label{fig:sources}
\end{figure}

Fig.~\ref{fig:sources} shows comparison of GWs sensitivity limits of optical cavities from Fig.~\ref{fig:sensitivity} with the fractional amplitude spectral densities of astrophysical sources that fall within the range of maximum resonance sensitivity. 
The plot shows predicted signals from  binary neutron star inspiral,
merger, and post-mergers, collapsing stellar cores, subsolar-mass BHs  mergers, and QCD axions and axion-like particles formed through BH superradiance.
While only 0.5~m, 1~m, and 2~m long cavities are presented in the graph,  the mechanical resonance position (the sensitivity peak) may be shifted by changing the cavity length ($f_0\sim1/L$).

{\it Coalescing neutron stars.}  
The predicted  gravitational-wave strain for a typical binary neutron star (NS) inspiral, merger,
and post-merger (NS-NS) is taken from~\cite{Ackley_2020_NEMO} and scaled to amplitude spectral density at the distance of 3 Mpc (size of the Local Group). Weaker signals from farther source distances form the shaded brick-red area in Fig.~\ref{fig:sources}. While the tidal effects emitted during the inspiral are outside possible sensitivity, the postmerger
signal above 1 kHz from massive neutron star remnants~\cite{Shibata_2006,Baiotti_2008} may be produced by majority of the binary NS mergers~\cite{Margalit_2019}. While matching the optical cavity mechanical resonance to the maximum of the GWs signal needs the cavity to be 2~m long, which can be technically challenging for the cryogenic SCS spacer, this kind of source has indisputable advantage of being already observed~\cite{Abbot_NS,Abbott_2019,Abbott_2020NS} making neutron star science low risk.

{\it Collapsing stellar cores.}
Theoretical predictions shows that  the process of a BH formation from fast spinning,
moderate-metallicity, massive stellar progenitors leads to seconds-long~\cite{OConnor_2011,de_Brye_2014} and high-amplitude GWs signal. 
The characteristic GWs spectra in the slow rotating model  is taken from~\cite{Cerda_2013_collapse}  and scaled to amplitude spectral density at the distance of 30 kpc (the Milky Way Galaxy size). Weaker signals from farther source distances form the shaded blue area in Fig.~\ref{fig:sources}.

{\it Coalescence of subsolar-mass BH binaries.} While there are no known
mechanisms through standard stellar evolution to produce sub-solar mass BHs, the observation of subsolar-mass BHs  merger will  be an indication of theirs primordial origin. This makes the potential observation particularly important since primordial
BHs  may contribute to the dark matter distribution~\cite{Zel'dovich_1967,Hawking_1971,Pani_2014} and  verify theories on dark matter triggered formation of BHs ~\cite{Shandera_2018,Singh_2021,Dasgupta_2021}. Potential GWs signals were calculated analytically for the innermost stable circular orbits of the BHs binaries with two equal masses, each $<$1~M$_{\odot}$, at the distance of   1~kpc \cite{Creighton}
(shaded green area in Fig.~\ref{fig:sources}).

{\it Axions and ALPs (axion-like particles) superradiance.} Light bosonic fields such as axions or ALPs can form gravitational bound states around a black hole~\cite{Damour_1976,Ternov_1978,Zouros_1979,Detweiler_1980}. Their occupation number grows exponentially at the cost of the
angular momentum and energy of the rotating BH through superradiance~\cite{Brito_2020} forming a coherent axion or ALP bound state emitting
gravitational waves~\cite{Arvanitaki_2010,Arvanitaki_2011,Yoshino_2014,Arvanitaki_2015,Brito_2015, Arvanitaki_2017, Brito_2017,Brito_2017b,Baumann_2019, Isi_2019,Isi_2020,Ng_2021,Aggarwal_2022}. Gravitational signals are expected to be
produced during  axions/ALPs transition between gravitationally bound levels, axions/ALPs annihilation to gravitons, and  bosenova collapse of the axion/ALPs cloud. The first two mechanisms should yield long lasting, monochromatic gravitational wave signals, since axions/ALPs involved in transitions and annihilations are in exact energy eigenstates of the BH potential.
The potential signal from axions/ALPs were calculated with the analytic
approximation from \cite{Isi_2019,Isi_2020} for the values used in  \cite{Aggarwal_2022}, i. e. for signals due to GWs produced from axions/ALPs around a BH in our galaxy within 10 kpc for $10^6$~s coherent integration time. The BH with initial masses of 1, 2 and 3~M$_{\odot}$ and initial spin of 0.9 were calculated for the dominant level ($l=m_l=1, n=0$) (solid lines in Fig.~\ref{fig:sources}).

\section{Conclusion}

In this paper we consider table-top ultra-stable optical cavity made with the most advanced present-day technologies and report that it can be used as a resonant-mass gravitational wave detector in the  2-20 kHz range of GWs spectrum. 
Moreover, despite the resonance character of the sensitivity, contrary to the metallic Weber bar detectors, the detection scheme allows observing potential GW also outside the resonance, although with smaller sensitivity.
We show that it allows detecting not only predicted GW signals from
such sources as binary neutron star mergers and post-mergers, subsolar-mass primordial black-hole
mergers, and collapsing stellar cores, but can reach new physics beyond standard model looking for
ultralight bosons such as QCD axions and axion-like particles formed through black hole superradiance.

\section*{Acknowledgements}
This project (20FUN08 NEXTLASERS) has received
funding from the EMPIR programme co-financed by
the Participating States and from the European Union’s
Horizon 2020 research and innovation programme.
The research is a part of the program
of the National Laboratory FAMO (KL FAMO) in Toru\'n,
Poland, and is supported by a subsidy from the Polish
Ministry of Science and Higher Education.

\bibliography{reference}

%\bibliographymet{reference}

\end{document}